\documentclass[
 reprint,
 amsmath,amssymb,
 aps,prl
]{revtex4-2}

\usepackage{graphicx}
\usepackage{dcolumn}
\usepackage{bm}
\usepackage{xcolor}

\renewcommand{\phi}{\varphi}
\newcommand{\ph}{\varphi}
\newcommand{\eps}{\varepsilon}
\renewcommand{\th}{\theta}

\renewcommand{\r}{\rho}

\renewcommand{\vec}[1]{\boldsymbol{\mathrm{#1}}}
\newcommand{\iin}{\mathrm{in}}
\newcommand{\out}{\mathrm{out}}

\newcommand{\rot}{\nabla\times }
\renewcommand{\div}{\nabla\cdot }
\newcommand{\grad}{\nabla}

\begin{document}

\preprint{APS/123-QED}

\title{Bound States in the Continuum in Compact Acoustic Resonators}% Force line breaks with \\

\author{Ilya Deriy}
 \email{ilya.deriy@metalab.itmo.ru}
%  \altaffiliation[Also at ]{Physics Department, XYZ University.}%Lines break automatically or can be forced with \\

\author{Ivan Toftul}%
\author{Mihail Petrov}
\author{Andrey Bogdanov}%
 \email{a.bogdanov@metalab.itmo.ru}
\affiliation{%
 Department of Physics and Engineering, ITMO University, 191002, St. Petersburg, Russia}
\date{\today}% It is always \today, today,

\begin{abstract}
We reveal that finite-size solid acoustic resonators can support genuine bound states in the continuum (BICs) completely localized inside the resonator. The developed theory provides the multipole classification of such BICs in the resonators of various shapes. It is shown how  breaking of the resonator's symmetry turns BICs into quasi-BICs manifesting themselves in the scattering spectra as high-Q Fano resonances. We believe that the revealed novel states will push the performance limits of acoustic devices and will serve as high-Q building blocks for acoustic sensors, antennas, and topological acoustic structures.     
\end{abstract}

\maketitle

%\section{Introduction \label{sec:intro}}

{\it Bound states in the continuum} (BICs) are the non-radiating states of an open system with a spectrum embedded in the continuum  of the radiating modes of the surrounding space~\cite{hsu2016bound}.  BICs were firstly predicted in quantum mechanics by von Neumann and Wigner in 1929~\cite{Neuman1929bound} but shortly after they were extended to the wave equations in general as their specific solutions. As a result, BICs were found in various fields of physics such as  atomic physics, hydrodynamics, and acoustics~\cite{fonda1963bound,ursell1951trapping,parker1966resonance}.  The zero radiative losses leads to diverging radiative quality factor (Q-factor) making BICs extremely prospective for the energy localisation  and enhancement of the incident fields. Since recently, these unique properties of BICs have been actively utilized in photonics, where they have already proven themselves as an effective platform for lasing, polaritonic, sensing, and optical harmonic generation applications~\cite{azzam2021photonic,koshelev2019nonradiating,kodigala2017lasing,koshelev2020subwavelength,kravtsov2020nonlinear,tittl2018imaging}.          

Decoupling the the resonance from all open scattering channels one can obtain a genuine BIC, which becomes  possible only if the number of the adjusting parameters is more than the number of the scattering channels. A typical example of a system with the finite number scattering channels is a resonator coupled to one or several waveguide modes~\cite{lepetit2014controlling,pilipchuk2020bound,lyapina2015bound}. Alternatively,  a finite number of the scattering channels may exist in infinite periodic structures~\cite{hsu2013observation,sadrieva2019experimental}. For finite size structures the number of scattering channels is infinite, and the existence of BICs in such systems is prohibited by {\it non-existence theorem}~\cite{hsu2016bound}. The only exception is the structures surrounded by a completely opaque shell providing decoupling of the internal resonances from the outside radiation continuum, which in quantum mechanics corresponds to an infinite high potential barriers, in acoustics to hard-wall boundaries, and in optics to perfect conducting walls or epsilon-near-zero barriers~\cite{monticone2014embedded,liberal2016nonradiating}. Finding of a genuine BIC in compact systems is a challenging fundamental problem, and its solution would make possible the implementation of subwavelength high-Q resonators having broad range of potential applications.

%%% figure 1 %%%%%%%%%%%%%%%%%%%%
\begin{figure}[b]
\includegraphics[width = 0.99\linewidth]{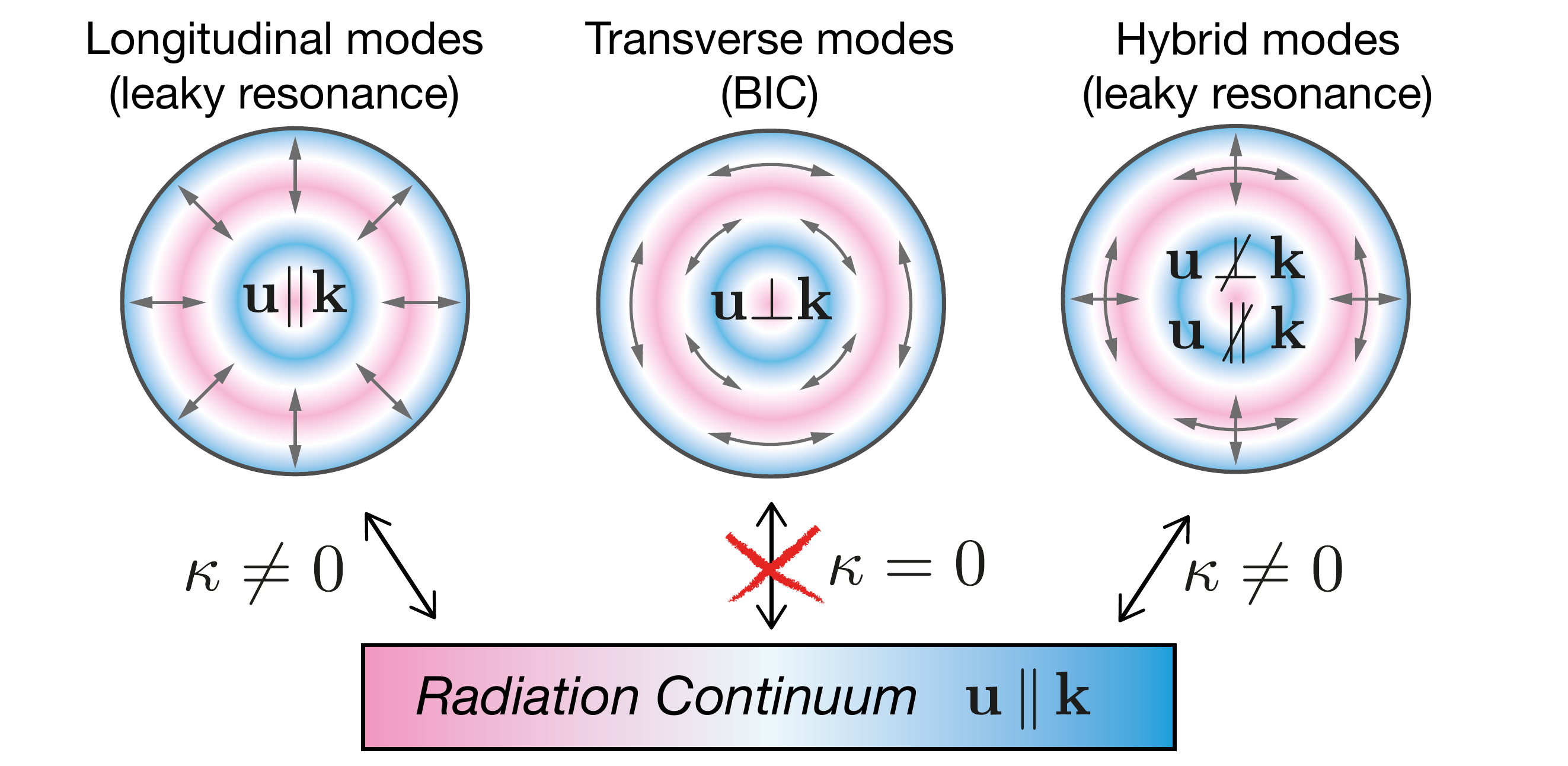}% Here is how to import EPS art
\caption{\label{fig:main_idea}
Polarization of eigenmodes in a solid acoustic resonator. 
The transversal modes ($\mathbf{u}\bot\mathbf{k}$) do not coupled to the radiation continuum forming a bound state in the continuum. Here, $\kappa$ is the coupling coefficient, $\mathbf{u}$ is the displacement vector, and $\mathbf{k}$ is the wave vector.
}
\end{figure}
%%%%%%%%%%%%%%%%%%%%%%%%%%%%%%%%%

% figure 2 %%%%%%%%%%%%%%%%%%%%%%%%%%%%%%%%%%%%%%%%%%
\begin{figure}[t]
\includegraphics[width=0.99\linewidth]{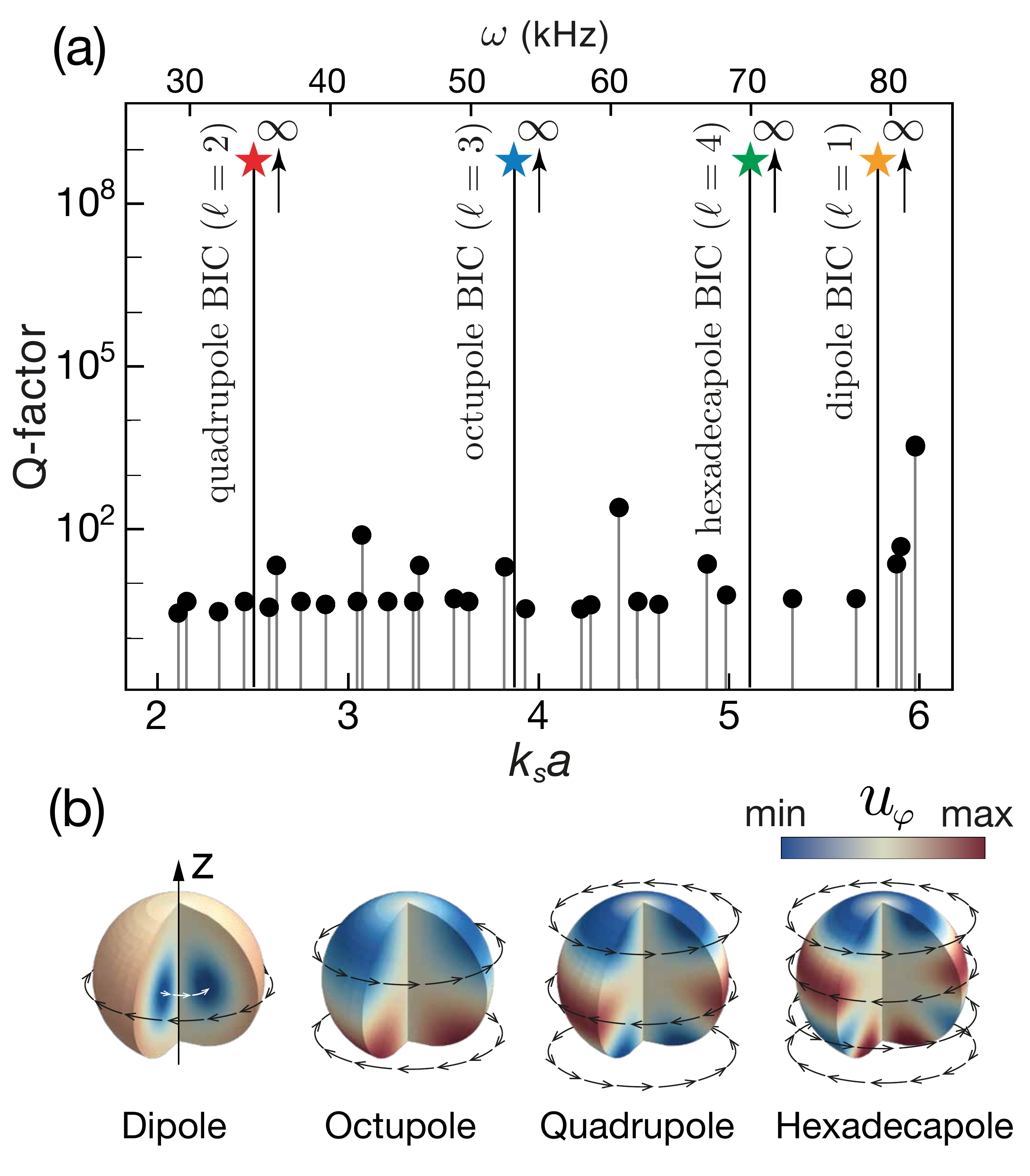}
\caption{\label{fig:modes}
Polarization-protected BICs in a solid spherical resonator. (a) Spectrum and radiative Q-factor of the resonator. Parameters of the resonator are mentioned in the text. (b) Distributions of the azimuthal component of the displacement ($u_\varphi$) for the dipole, quadrupole, octupole, and hexadecapole BICs with $m=0$.}
\end{figure}
%%%%%%%%%%%%%%%%%%%%%%%%%%%%%%%%%%%%%%%%%%%%%%%%%%%%%

In this letter, a backdoor in the 'non-existence theorem' is revealed. While BICs are prohibited in finite photonic and quantum mechanical systems, we show that acoustics is relieved of these constraints. We propose a genuine acoustic BIC in compact solid resonators placed in nonviscous fluids (gas or liquid). The origin of the unexplored BICs is illustrated in Fig.~\ref{fig:main_idea}. In acoustics, in contrast to photonics and quantum mechanics, there are two types of waves: (i) {\it pressure waves} with longitudinal polarization ($\mathbf{u}\|\mathbf{k}$) and (ii) {\it shear waves} with transversal polarization ($\mathbf{u}\bot\mathbf{k}$)~\cite{kinsler1999fundamentals}. Here, $\mathbf{u}$ is the displacement vector and $\mathbf{k}$ is the wave vector. While in solids, both waves coexist, nonviscous fluids host only pressure waves which are longitudinal. In general case of an arbitrary shaped  solid resonator, all the eigenmodes are hybrid containing  longitudinal and transverse components  hybrid and, thus, are coupled to the radiation continuum. However, one can find specific shapes of the resonators allowing purely torsion modes  which displacement is tangential at each point of the resonator boundary. The torsion oscillations do not produce pressure on the surrounding fluid being completely decoupled from the radiation continuum. Their  energy remains perfectly confined inside the resonator forming ideal genuine acoustic BICs.

We may designate the proposed non-radiating states as {\it polarization-protected acoustic BICs} as they are possible exclusively due to the fact that non-viscous fluids support  longitudinal waves only, while solids host both longitudinal and transversal waves. Due to  similar reasons,  the shear modes in a solid slab appear to be non-radiating~\cite{quotane2018trapped}. To be frank, BICs in compact resonators can exist even in optics, for example, in the form of the radial plasmonic oscillations in a metal spherical particle at the plasma frequency. In this case, the longitudinal plasmonic oscillations are not coupled to the far-field electromagentic radiation which is purely transversal. However, the observation of these modes is hindered due to high losses in plasmonic nanostructures.

We start with considering a problem of eigen modes in a rigid sphere of radius $a$ in order to show rigorously that the polarization-protected BICs exist in compact resonators. We assume that the  sphere is made of a solid isotropic material surrounded by gas or fluid environment. The displacement field $\mathbf u(\mathbf r)$ inside and outside the sphere obeys the Helmholtz equation~\cite{kinsler1999fundamentals} 
\begin{equation}
    \Delta \mathbf u_i^j(\vec r) + (k_i^j)^2\vec u_i(\vec r) = 0.
    \label{eq:Helmholtz}
\end{equation}
Here, the lower index $i = s,p$ encodes the displacements of shear or pressure waves respectively, the upper index $j=\{\iin, \out\}$ corresponds to the fields inside and outside of the resonator respectively, $k_i^j = \omega/c_i^j$ is the wavevector, and $c_i^j$ is the velocity of displacement waves.
The boundary conditions at the surface of a solid sphere placed in gas or fluid can be written  in spherical coordinates ($r,\theta,\varphi$) as follows \cite{isakovich1973general}: 
\begin{equation}
    \sigma_{rr}^\iin = -p^\out, \:\: \sigma_{r\theta}^\iin = 0,\:\: \sigma_{r\varphi}^\iin = 0,\:\: u_r^\iin = u_r^\out,
    \label{eq:boundary_equations}
\end{equation}
where $\sigma_{ij} = \lambda\delta_{ij}\operatorname{Tr}\hat \varepsilon +\mu \varepsilon_{ij}$ is the Cauchy stress tensor, $2\hat\varepsilon = [ \nabla\mathbf u +\left(\nabla\mathbf u\right)^\mathrm{T} ]$ is the strain tensor, $p^\out$ is the gas pressure, which is connected to the displacement field as $\rho^\out \omega^2\mathbf u^\out = \nabla p^\out$ (see Supplemental Material for details). We also assume that outside the sphere the solution has a form of the outgoing waves. Based on that, the solutions of the vector Helmholtz equation\eqref{eq:Helmholtz} can be written in terms of the vector spherical harmonics  \cite{bohren2008absorption} 
\begin{multline}
    \vec u^j(\vec r) = \sum_{\ell m}a_{\ell m}^j\vec M_{\ell m}(\vec r, k_s^j)+
    \\
    +b_{\ell m}^j\vec N(\vec r, k_s^j)+ c_{\ell m}^j\vec L_{\ell m}(\vec r, k_p^j).
    \label{eq:vsh_decomposition}
\end{multline}
Here, $\ell = 0,1,2,...$ is the total angular momentum quantum number and $m = 0,\pm1,...,\pm\ell$ is the projection of the angular momentum on the $z$-axis (magnetic quantum number) [see Fig.~\ref{fig:modes}(b)]. While outside the cavity one should put  $a_{\ell m}^\out = b_{\ell m}^\out=0$ for all $\ell $ and $m$ as fluid(gas) supports only longitudinal waves, inside the cavity one should account for all the vector harmonics.
%%% figure 3 %%%%%%%%%%%%%%%%%%%%
\begin{figure}[t]
\includegraphics[width = 0.99\linewidth]{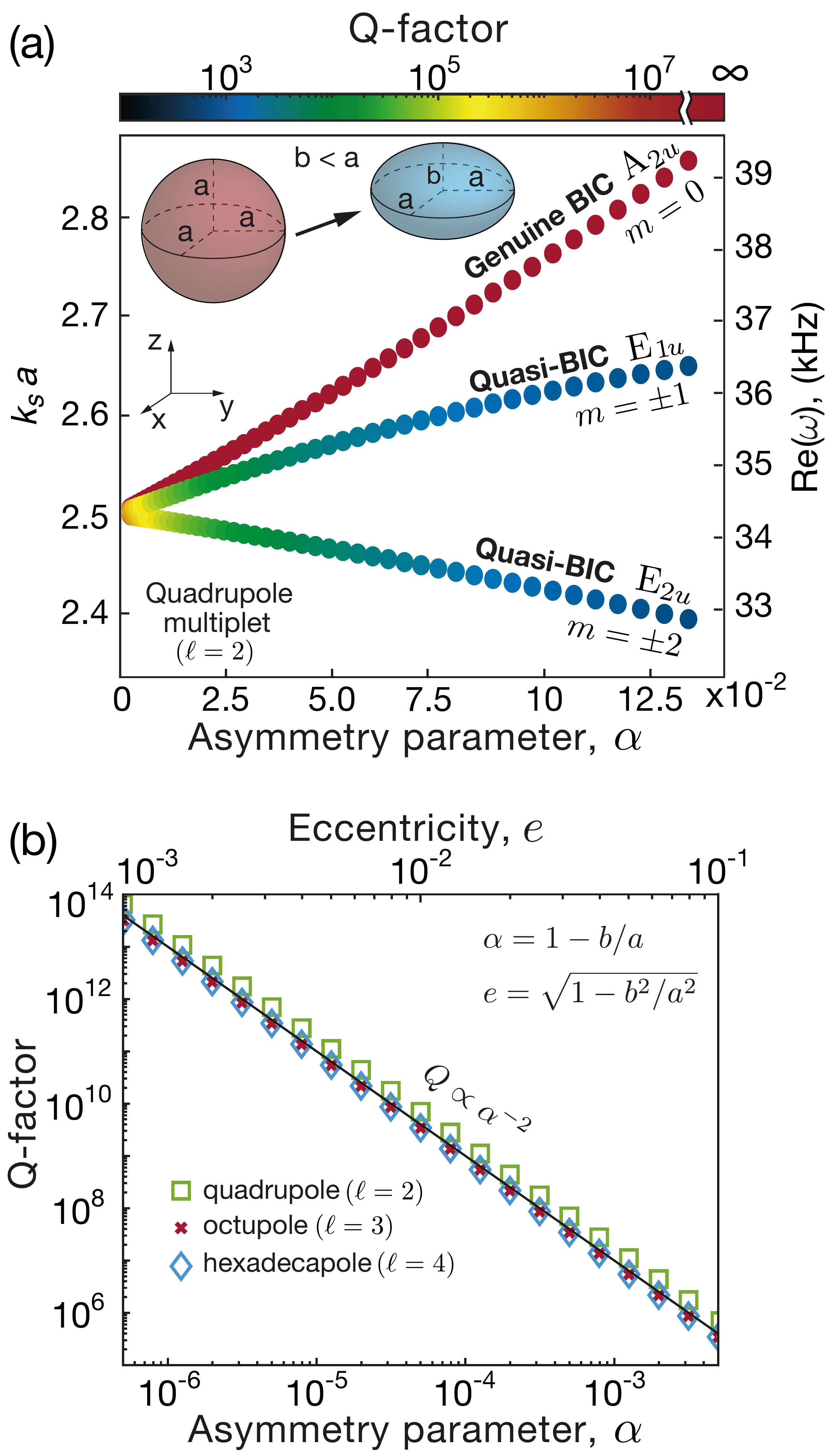}%
\caption{\label{fig:excentricitet}
Splitting of a quadrupole multiplet into BIC and quasi-BIC in a solid spheroid. (a) Frequency shift and Q-factors versus asymmetry parameter $\alpha=1-b/a$. (b) Dependence of the Q-factor on the asymmetry parameter $\alpha$ for quadrupole, octupole, and hexadecapole quasi-BICs with $m=\pm2$.}
\end{figure}
%%%%%%%%%%%%%%%%%%%%%%%%%%%%%%%%%
As a result, the homogeneous system of equations on the expansion coefficients can be obtained
\begin{equation}
    \hat {\vec D}_{\ell m} \mathbf f_{\ell m} = 0,
    \label{eq:matrix}
\end{equation}
where vector $\vec f_{\ell m} = \left[ a^\iin_{\ell m}, b^\iin_{\ell m}, c^\iin_{\ell m}, c^\out_{\ell m} \right]^\mathrm{T}$ is the vector of coefficients, and  $\hat {\vec D}_{\ell m} = \operatorname{diag}[\hat {\vec D}_{\ell m}^{1\times1},\:  \hat {\vec D}_{\ell m}^{3\times3}]$ is a block-diagonal matrix with the explicit form provided in Supplemental Material. In virtue of the block-diagonal form of $\hat {\vec D}_{\ell m}$, the equation on eigenfrequencies is factorized:
\begin{equation}
    \underbrace{\operatorname{det}\hat {\vec D}_{\ell m}^{1\times1}}_{\text{BIC}}\cdot\underbrace{\operatorname{det}\hat{\vec D}_{\ell m}^{3\times3}}_{\text{Rad. modes}}=0.
    \label{eq:common_eigenfrequency}
\end{equation}
Generally, the eigen frequency can be rescalled by the factor of $c_s/a$. However, to have an illustrative physical example we will use the secondary  frequency axis corresponding to the sphere of radius $a=5$~cm  placed in air with density $\rho_0 = 1.23$ kg/$\mathrm{m}^3$. The density of the material of the sphere equals to $\rho = 10\rho_0$, while the velocity of shear and pressure waves  are taken equal to $c_s = 2c_0$, $c_p = 3c_0$, respectively, where $c_0 = 343$ m/s is the velocity of sound in the air.  

The spectrum of the resonator (eigenfrequencies and Q-factors) obtained from numerical solution of  Eq.~\eqref{eq:common_eigenfrequency} is shown in Fig.~\ref{fig:modes}(a).  It consist of radiative modes and BICs with infinitely high radiative Q-factors. One can show that purely real eigenfrequencies corresponding to BICs satisfy the equation  $\operatorname{det}\hat{\vec D}_{\ell m}^{1\times1}=0$, which  can be written explicitly as follows (see Supplemental Material for details):
\begin{equation}
    (1-\ell)j_\ell(x)+x j_{\ell+1}(x) = 0.
    \label{eq:m_eigenfrequency}
\end{equation}

\noindent Here $j_\ell(x)$ it the spherical Bessel function, and $x = k_sa$. Due to spherical symmetry of the problem, the obtained equation does not depend on $m$, therefore, the its solution is  $(2\ell+1)$-fold degenerate. The table containing the roots of Eq.~\eqref{eq:m_eigenfrequency} is provided in Supplemental Material.   

It also follows from Eqs.~\eqref{eq:matrix} and \eqref{eq:common_eigenfrequency} that  $b^\iin_{\ell m}=c^\iin_{\ell m}=c^\out_{\ell m}=0$ for BICs and, consequently,  the displacement field $\mathbf{u}$ contains  only  vector harmonics $\vec M_{\ell m}$ and they are completely localized inside the resonator. Indeed, $\vec M_{\ell m}$ harmonics correspond to the torsion oscillations which totally lack of radial components in sharp contrast to $\vec L_{\ell m}$ and $\vec N_{\ell m}$ and, thus,  can not excite pressure waves in the surrounding fluids. The distribution of the displacement field for BICs with $m=0$ and $\ell=1,2,3,4$ is shown in Fig.~\ref{fig:modes}(b). In the light of the modes structure, we should note that  Eq.~\eqref{eq:m_eigenfrequency} can be also obtained from Eq.~\eqref{eq:Helmholtz} applying stress-free boundary conditions~\cite{tamura1982lattice,tamim2019elastic}.

A curios fact deserving special attention is that the fundamental acoustic BIC in a solid sphere is a quadrupole mode ($\ell=2$) rather than a dipole. This fact can be understood intuitively: the time-average angular momentum of the resonator should be zero, i.e. the resonator should not rotate as a whole. For the dipole modes, the external and internal layers oscillate in anti-phase compensating the rotation [see Fig.~\ref{fig:modes}(b)], and it results in a quite large radial wavenumber $k_sa\approx5.8$. For the quadrupole mode $k_sa\approx2.5$ and the oscillation phase does not change along the radial direction. Thus, the time-average angular momentum is compensated by anti-phase oscillations of the upper and lower hemispheres [see Fig.~\ref{fig:modes}(b)]. 

A BIC according to its definition is completely decoupled from all propagating waves of the surrounding space, and, thus, it cannot be excited from the far-field by pressure waves. However, the excitation is possible by near-field sources or due to nonlinear effects~\cite{bulgakov2015all,chukhrov2021excitation,yuan2020excitation}. Another efficient method, that is the most used in practice, is based on introduction of small coupling between the BIC and radiative modes. Therefore, a genuine BIC turns into a {\it quasi-BIC} (q-BIC) -- a high-Q state that manifests itself in the scattering spectrum as a narrow Fano resonance~\cite{koshelev2018asymmetric}. Recently, the q-BICs were suggested as very promising candidates for sensing, lasing, and nonlinear optics applications~\cite{koshelev2019meta, koshelev2019nonlinear,tittl2018imaging,liu2019high,leitis2019angle,vaskin2019light,liu2018extreme}.              

In order to show how a genuine acoustic BIC turns into a q-BIC, we slightly deform the spherical resonator of radius $a$ into an oblate spheroid with the semi-axes $a$ and $b$ [see inset in Fig.~\ref{fig:excentricitet}(a)]. The dependence of the resonant frequency and Q-factor for the quadrupole BIC ($\ell=2$) on the asymmetry parameter $\alpha=1-a/b$ is shown in Fig.~\ref{fig:excentricitet}(a). As it was mentioned above, the BICs in spherical resonator are $(2\ell+1)$-fold degenerate multiplets. In the spheroid, this multiplet splits into one singlet state corresponding to the BIC and $2\ell$ of two-fold degenerate doublet q-BICs (see Supplemental Material for details) [Fig.~\ref{fig:excentricitet}(a)]. The Q-factor of the q-BICs drops quadratically with the asymmetry parameter $\alpha$ [see Fig.~\ref{fig:excentricitet}(b)] that completely agrees with the general theory of q-BICs~\cite{koshelev2018asymmetric}. 

In terms of the group theory, the quadrupole BIC in a spheroid ($D_{\infty h}$ symmetry) corresponds to one-dimensional irreducible representation (irrep) $A_{2u}$, and the doublets of q-BICs correspond to two-dimensional irreps $E_{1u}$ and $E_{2u}$. The degeneracy of the q-BICs remains due to the rotational symmetry of the spheroid, thus, the doublet can be associated with clockwise- and counterclockwise rotational modes. Following Ref.~\cite{gladyshev2020symmetry} we can write the multipole series of the BICs and q-BICs, and selection rules for their excitation by a plane pressure wave incident from different directions (see Table~\ref{tab:quadrupole_BIC}). The survivor BICs in a spheroid are contributed only by the vector harmonics $\mathbf{M}_{\ell m}$ with even $\ell$ and $m=0$. As a spheroid has a symmetry plane ($xy$), there is a second series of genuine BICs contributed by the vector harmonics $\vec M_{\ell m}$ with odd $\ell$ and $m=0$ (see Supplemental Material for details).

One may see from Table~\ref{tab:quadrupole_BIC} that q-BICs from $E_{1u}$ can be excited by a plane pressure wave propagating along the $x$- or $y$-axis. The q-BICs from $E_{2u}$ can be excited only at oblique incidence. Indeed, the excitation from the $x$-, $y$- directions is forbidden due to inconsistency between the parity of the mode and the incident wave, and the excitation from the $z$-direction is forbidden due to the fact that the incident wave contains only the harmonics with $m=0$.

%%% figure 4 %%%%%%%%%%%%%%%%%%%%
\begin{figure}[t]
\includegraphics[width = 1.00\linewidth]{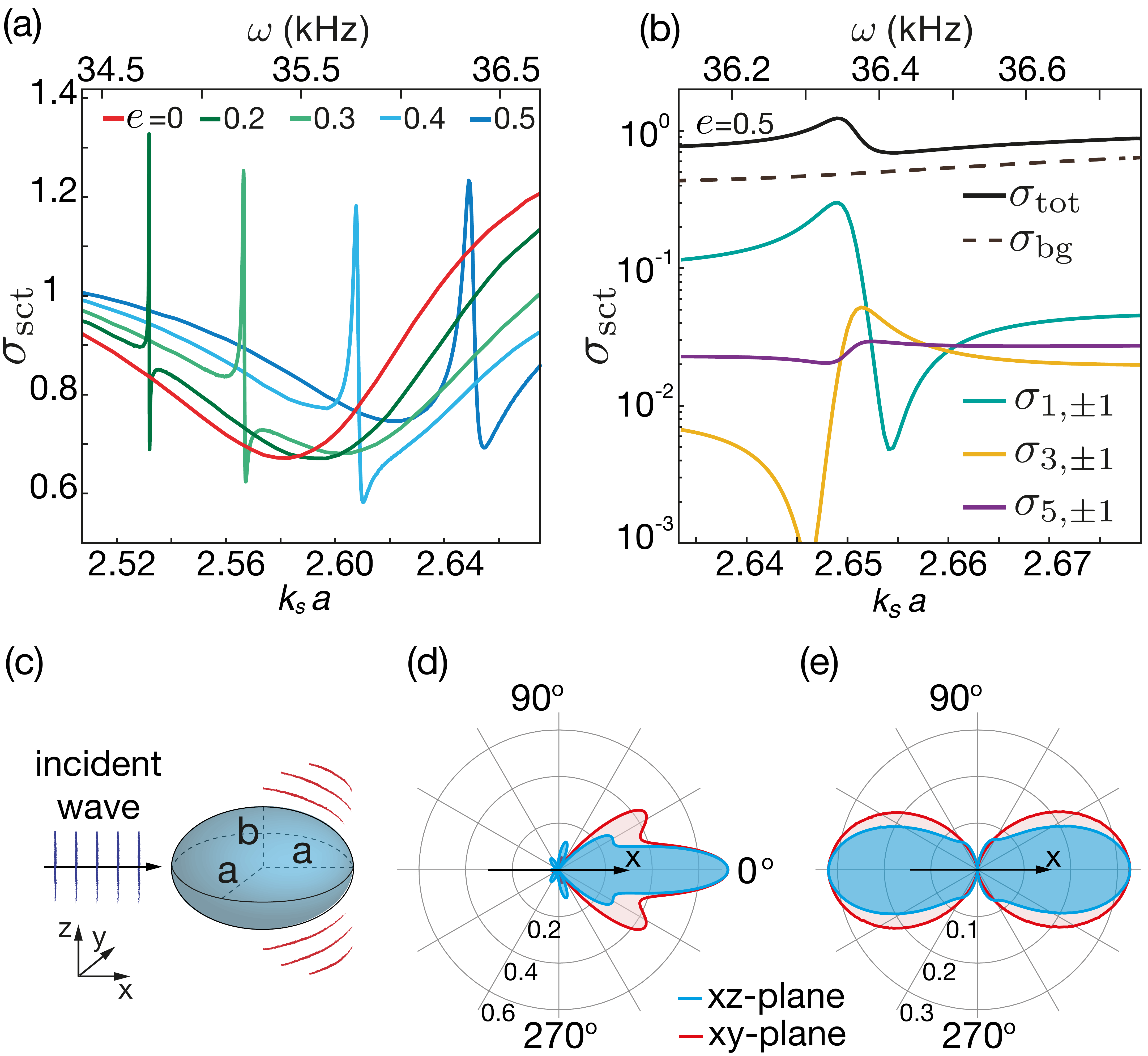}%
\caption{\label{fig:scattering}
(a) Spectrum of scattering efficiency of a solid spheroid calculated for various eccentricities $e$. (b) Spectrum of the total and partial scattering efficiencies calculated near q-BIC ($E_{1u}$) for $e=0.5$. (c) The excitation scheme. (d) Directivity patterns of the total scattered field and (d) scattered field accounted for   the resonant contribution of q-BIC only. The diagrams are plotted for $e=0.1$ at the resonant frequency of q-BIC  ($k_sa=2.51$).}
\end{figure}
%%%%%%%%%%%%%%%%%%%%%%%%%%%%%%%%% 

Figure~\ref{fig:scattering}(a) shows the scattering efficiency $\sigma_{\text{sct}}$ (scattering cross-section normalized to the geometric cross-section) of a solid spheroid excited by a plane pressure wave propagating along the $x$-axis. The scheme of excitation is  illustrated in Fig.~\ref{fig:scattering}(c). The spectra for different eccentricities $e$ were calculated numerically using COMSOL Multiphysics. One can see that q-BIC corresponding to irrep $E_{1u}$ appears as a high-Q Fano resonance that collapses when $e$ tends to zero manifesting the formation of a genuine BIC.  Figure~\ref{fig:scattering}(b) shows the contribution of the resonant and non-resonant scattering channels $\sigma_{\ell m}$ to the total scattering efficiency $\sigma_\text{tot}$. Figures~\ref{fig:scattering}(d) and \ref{fig:scattering}(e) show the directivity patterns for the total scattered field and scattered field accounted for only the resonant harmonics of q-BIC (see Table~\ref{tab:quadrupole_BIC}). The diagrams are plotted for the speroid with $e=0.1$ at the resonant frequency of q-BIC  ($k_sa=2.51$). One can see that the q-BIC behaves as dipole in the far-filed but the non-resonant scattering drastically changes  the directivity pattern.

As we see from Fig.~\ref{fig:excentricitet}(a), the polarization-protected acoustic BICs exist not only in a sphere but also in a spheroid. A reasonable question is {\it 'Can we find all possible shapes of resonators capable of supporting such BICs?'}. To answer this question one can address to group theory. Table~\ref{tab:quadrupole_BIC} shows that the symmetry breaking results in mixing multipoles and each eigenmode represents an infinite series of multipoles. Thus, in the general case of an arbitrary shape resonator, all the modes are radiative. However, the multipole mixing occurs according to the selection rules defined by the symmetry group of the resonator. BICs will survive under reduction of resonator's symmetry if the multipoles $\vec M_{\ell m}$ do not couple with $\vec N_{\ell m}$ and $\vec L_{\ell m}$. Thus, BICs are allowed only if the symmetry group of the resonator contains the irreducible representations with a basis forming by the harmonics $\mathbf{M}_{\ell m}$ only. One can show that such a requirement is fulfilled only for the symmetry groups containing the infinite-fold rotation axis. These are $D_{\infty h}$ (cylinder, spheroid, dimer or linear chain of identical equidistant spheres or cylinders) and $D_{\infty v}$ (cone or an arbitrary body of revolution without horizontal symmetry plane).

% table 2 %%%%%%%%%%%%%%%%%%%%%%%%%%%%%%%%%%%
\begin{table}[t]
\caption{\label{tab:quadrupole_BIC} 
Splitting of a quadrupole multiplet BIC ($\ell=2$) in a spheroid. Multipole content of the modes and selection rules for their excitation by a plane wave.
}
\begin{ruledtabular}
\begin{tabular}{ccccc}
Irrep&
Mode&   
\text{Multipoles}&
\begin{tabular}[c]{@{}c@{}}Excitation\\ along $x$ or $y$ \end{tabular}&
\begin{tabular}[c]{@{}c@{}}Excitation\\ along $z$ \end{tabular}
\\
%\colrule
%A_{1g} & BIC & $\mathbf{M}_{1,0}$,...,$\mathbf{M}_{2s+1,0}$ & No & No \\
\colrule
$A_{2u}$ & BIC & $\mathbf{M}_{2,0}$,...,$\mathbf{M}_{2s,0}$ & No & No \\
\colrule
$E_{1u}$ & q-BIC & \begin{tabular}[c]{@{}l@{}}$\mathbf{L}_{1,\pm1}$...,$\mathbf{L}_{2s+1,\pm1}$\\ $\mathbf{N}_{1,\pm1}$...,$\mathbf{N}_{2s+1,\pm1}$ \\ 
$\mathbf{M}_{2,\pm1}$...,$\mathbf{M}_{2s,\pm1}$
\end{tabular}  & Yes & No  \\
\colrule
$E_{2u}$ & q-BIC &\begin{tabular}[c]{@{}l@{}}
$\mathbf{L}_{3,\pm2}$,...,$\mathbf{L}_{2s+1,\pm2}$\\
$\mathbf{N}_{3,\pm2}$,...,$\mathbf{N}_{2s+1,\pm2}$ \\
$\mathbf{M}_{2,\pm2}$,...,$\mathbf{M}_{2s,\pm2}$
\end{tabular} & No & No \\
\end{tabular}
\end{ruledtabular}
\end{table}
%%%%%%%%%%%%%%%%%%%%%%%%%%%%%%%%%%%%%%%%%%%%%%
In this prospective, a special attention should be paid to the polarization-protected acoustic BICs in a solid cylinder of radius $a$ and height $h$. Though the eigenvalue problem for an open cylindrical resonator cannot be solved analytically since the variables can not be separated, it becomes possible for particular BIC solution. Indeed, BICs are perfectly localized inside the resonator, thus, the system can be considered as a closed one. Therefore, the eigenfrequencies can be calculated analytically even in the cylindrical geometry using the stress-free boundary conditions:
$\left({\omega}/{c_s}\right)^2 =\left({\alpha_n}/{a}\right)^2+\left({\pi q}/{h}\right)^2$. Here $\alpha_n$ is the $n$'th root of the Bessel function $J_2(x)$, and $q$ is an integer.

The total Q-factor ($Q_{\text{tot}}$) of acoustic BICs is limited by the absorption Q-factor ($Q_{\text{abs}}$) that is defined by the attenuation of shear waves in real materials. For example, the longitudinal loss factor $\eta = E''/E'$ of steel is reported to be on the order of $10^{-4}-10^{-5}$ \cite{zhang1993documentation, irvine2004damping}, where $E = E'+iE''$ is the complex Young modulus. Loss factor of the same order is reported for  different ceramics and glasses \cite{zhang1993documentation}, while for silica the loss factor is on the order of $10^{-6}$. Since the longitudinal and shear loss factors are of the same order \cite{pritz2009relation}, one can expect the total Q-factor of BICs in real materials to be of the order of $Q=1/\eta \sim 10^{4}$ to $10^{6}$.
      
In conclusion, we have revealed that genuine acoustic bound states in the continuum may exist in compact solid resonators with a rotational symmetry placed in  gas or nonviscous fluid environment. The predicted states are possible due to polarization mismatch between the shear waves in  solid resonator and pressure waves in the surrounding media. We believe that our findings are an important step in the development of  high-Q resonant acoustics, and the revealed novel BICs in compact structures will serve as building blocks for acoustic antennas, high-sensitive acoustic sensors, and topological acoustic structures.    

%======acknowledgement=======

The authors thank K.~Frizyuk, K. Koshelev, and Yu. Kivshar for useful discussions and suggestions. The work is support by the Russian Science Foundation (project \#20-72-10141). A.\,B. acknowledges a support from the “BASIS” Foundation. 

% \nocite{*}

\providecommand{\noopsort}[1]{}\providecommand{\singleletter}[1]{#1}%
%

%%%%%%%%%% Merge with supplemental materials %%%%%%%%%%
\widetext
\clearpage
\begin{center}
\textbf{\large Supplemental Materials: \\ 
Bound States in the Continuum in Compact Acoustic Resonators}
\end{center}
%%%%%%%%%% Merge with supplemental materials %%%%%%%%%%
%%%%%%%%%% Prefix a "S" to all equations, figures, tables and reset the counter %%%%%%%%%%
\setcounter{equation}{0}
\setcounter{figure}{0}
\setcounter{table}{0}
\setcounter{page}{1}
\makeatletter
\renewcommand{\theequation}{S\arabic{equation}}
\renewcommand{\thefigure}{S\arabic{figure}}
\renewcommand{\bibnumfmt}[1]{[S#1]}
\renewcommand{\citenumfont}[1]{S#1}
\section{Wave equation for the displacement field in isotropic elastic media.}
\label{sec:wave_equation}
Equation of motion for the elastic media can be written as \cite{landauelasticity}
\begin{equation}
    \rho \ddot{\vec u} = \div \hat{\vec \sigma},
    \label{eq:eqmot}
\end{equation}
where $\rho$ is the density of the media, $\vec u$ is the displacement field and $\hat{\vec \sigma}$ is the Cauchy stress tensor.
Stress tensor is connected with displacement field via the Hooke's law
\begin{equation}
    \hat{\vec \sigma} = \hat{\mathrm C}:\hat{\vec \varepsilon},
    \label{eq:hooke_init}
\end{equation}
where $\hat{\mathrm{C}}$ is the fourth-rank tensor of the elastic constants and $\hat{\vec \varepsilon} = 1/2\left[\nabla\vec u + \left(\nabla\vec u\right)^\mathrm{T}\right]$ is the infinitesimal strain tensor.
In an isotropic elastic media, tensor of elastic constants is (in Voigt notation) \cite{nye1985physical}
\begin{equation}
    \hat{\mathrm{C}} =
    \begin{bmatrix}
    \lambda+2\mu & \lambda & \lambda & 0 & 0 & 0
    \\
    & \lambda + 2\mu & \lambda & 0 & 0 & 0
    \\
    & &\lambda + 2\mu & 0 & 0 & 0 
    \\
     & & & \mu & 0 & 0
    \\
    & & & &  \mu & 0
    \\
    & & & & &   \mu
    \end{bmatrix},
\end{equation}
or equivalently
\begin{equation}
    \mathrm C_{ijkl} = \lambda \delta_{ij}\delta_{kl} + \mu\left(\delta_{ik}\delta_{jl}+\delta_{jk}\delta_{il}\right),
    \label{eq:constants}
\end{equation}
thus giving the Hooke's law \eqref{eq:hooke_init} for isotropic media
\begin{equation}
    \sigma_{ij} = \lambda\delta_{ij}\eps_{kk} + 2\mu \eps_{ij}.
    \label{eq:hooke}
\end{equation}
Here $\lambda$ and $\mu$ are the Lame parameters.
By substituting Hooke's law (\ref{eq:hooke}) for isotropic media into the equation of motion (\ref{eq:eqmot}), and considering time-harmonic fields $\vec u(\vec r, t)\sim \vec A(\vec r) e^{-i\omega t}$, one can obtain
\begin{equation}
   \rho\omega^2\vec u + (\lambda+2\mu)\grad\div \vec u - \mu\rot\rot \vec u = 0.
   \label{eq:wveq_initial}
\end{equation}
Using the Helmholtz theorem, displacement field can be expanded into the curl-free $\vec u_s$ and divergence-free $\vec u_p$ components
\begin{equation}
\begin{aligned}
        \vec u = \vec u_{s} &+ \vec u_p,
        \\
        \div\vec u_s = 0, \: &\rot\vec u_p = 0.
\end{aligned}
    \label{eq:helmholtz_decompose}
\end{equation}
Substitution of the expansion Eq.~\eqref{eq:helmholtz_decompose} into the Eq.~(\ref{eq:wveq_initial}) with the help of the relation $\Delta = \grad\div - \rot\rot$ will split equation (\ref{eq:wveq_initial}) into two equations. One for the pressure (curl-free) waves and one for the shear (divergence-free) waves
\begin{equation}
    \begin{aligned}
        \Delta \vec u_i + k_i^2 \vec u_i = 0,
    \end{aligned}
    \label{eq:helmholtz}
\end{equation}
where index $i = p,s$ (pressure or shear), $k_i = \omega/c_i$ is the wavenumber of the pressure or shear wave, $c_p^2 = (\lambda+2\mu)/\rho$, $c_s^2 = \mu/\rho$ are the speed of the pressure or shear wave in the media. One can see that equation (\ref{eq:helmholtz}) is the Helmholtz equation.

%%% p and u %%%%%%%%%%%%%%%%%%%%%%%%%%%%%%%%%%%%%%%%%%%%%%%%%%%%%%%%%%%%%%%%%%%%%%
\section{Connection between pressure and displacement.}
\label{sec:pressure_displacement}

In the case of time-harmonic fields in nonviscious fluid (or gas) pressure $p(\mathbf r)$ and velocity $\vec v(\mathbf r)$ fields are connected as (in the first order of the perturbation theory) \cite{bruus2012acoustofluidics}
\begin{equation}
    \mathbf v(\mathbf r) = \dfrac{-i}{\rho_0\omega}\nabla p(\mathbf r),
\end{equation}
where $\rho_0$ is the density of the fluid.
On the other hand, velocity is the time-derivative of the displacement $\vec v (\vec r) = \dot{\vec u}(\vec r) = -i\omega \vec u(\vec r)$. Therefore,
\begin{equation}
    \vec u(\vec r) = \dfrac{1}{\rho_0\omega^2}\nabla p(\vec r).
    \label{eq:pu_connection}
\end{equation}

%%% Vector Harmonics %%%%%%%%%%%%%%%%%%%%%%%%%%%%%%%%%%%%%%%%%%%%%%%%%%%%%%%%%%%%%%%%%%%%%%
\section{Vector harmonics.}
\label{sec:vector_harms}
Vector harmonics are solutions of the vector Helmholtz equation and defined as follows \cite{bohren2008absorption}
\begin{equation}
   \vec L = \nabla \psi,\:\: \vec M = \nabla \times (\vec c \psi),\:\: \vec N = \dfrac{1}{k}\nabla\times \vec M,
   \label{eq:VH_Def}
\end{equation}
where $\psi$ is the solution of the scalar Helmholtz equation and $\vec c$ is called direction vector.
Since
\stepcounter{equation}
\begin{enumerate}
    \item $\vec M$ and $\vec N$ are the solutions of the Helmholtz equation
    \begin{equation}
        \rot\vec N = \dfrac{1}{k}\rot\rot \vec M = k\vec M,
        \tag{\theequation a}
        \label{eq:vhprop_a}
    \end{equation}
    \item $\vec M$ and $\vec N$ are solenoidal,
    \begin{equation}
        \div\vec M = \div\vec N = 0,
        \tag{\theequation b}
        \label{eq:vhprop_b}
    \end{equation}
    \item $\vec L$ is potential,
    \begin{equation}
        \rot\vec L = 0.
        \tag{\theequation c}
        \label{eq:vhprop_c}
    \end{equation}
\end{enumerate}

\subsection{Spherical vector harmonics}
In spherical system of coordinates  $(r, \theta, \phi)$ solution of the scalar Helmholtz equation can be written as
\begin{equation}
    \psi_{\ell m}(\vec r, \vec k) = z_\ell(k r)P_\ell^m(\cos\theta)e^{im\phi},
    \label{eq:solution_helmholtz_sph}
\end{equation}
where $z_\ell(x)$ is the spherical Bessel function of any kind. If one will take radius vector $\vec r$ as a direction vector, one will get explicit view of vector spherical harmonics
\begin{equation}
\begin{aligned}
    &\vec L_{\ell m} \left(\vec r, \vec k\right)= \left[\hat{\vec e}_r \dfrac{\partial z_\ell(k r)}{\partial r}P_\ell^m(\cos\th) + \hat{\vec e}_\theta \dfrac{z_\ell(k r)}{r}\dfrac{\partial P_\ell^m(\cos\th)}{\partial \theta}
    +\hat{\vec e}_\phi\dfrac{i m }{r \sin\theta}z_\ell(k r) P_\ell^m(\cos\th)\right]e^{i m \phi}
    \\
    &\vec M_{\ell m} \left(\vec r, \vec k\right) = \left[\hat{\vec e}_\theta \dfrac{i m }{\sin\theta}z_\ell(k r) P_\ell^m(\cos\th) - \hat{\vec e}_\phi z_\ell(k r)\dfrac{\partial P_\ell^m(\cos\th)}{\partial \theta}\right]e^{i m \phi}
    \\
    &\vec{N}_{\ell m} \left(\vec r, \vec k\right) = \left[\hat{\vec e}_r\frac{z_\ell(kr)}{k r} n(n+1) P_\ell^{m}(\cos\theta)+{\hat{\vec e}_\theta}\frac{1}{k r} \frac{\partial\left(r z_\ell(kr)\right)}{\partial r} \frac{\partial P_\ell^{m}(\cos\theta)}{\partial \theta}+ \hat{\vec{e}}_{\phi}\frac{1}{k r} \frac{\partial\left(r z_\ell(kr)\right)}{\partial r} \frac{i m}{\sin \theta} P_\ell^{m}(\cos\theta)\right]e^{im\phi}
\end{aligned}
\end{equation}

\subsection{Cylindrical vector harmonics}
In cylindrical system of coordinates $(\rho, \phi, z)$ solution of the scalar Helmholtz equation is
\begin{equation}
    \psi_m (\vec r, \vec k)= Z_m(k_\rho\rho)e^{im\phi}e^{ik_z z},
    \label{eq:solution_helmholtz_cyl}
\end{equation}
where $Z_m(x)$ is the Bessel function of any kind.
By taking the direction vector $\vec c = \hat{\vec e}_z$ and following definitions from the equation (\ref{eq:VH_Def}), one can obtain explicit view of vector cylindrical harmonics
\begin{equation}
    \begin{aligned}
&\mathbf{L}_{m}\left(\vec r, \vec k\right)=\left[
\hat{\boldsymbol{e}}_\rho\dfrac{\partial Z_m(k_\rho\rho)}{\partial \rho} + 
\hat{\vec e}_\varphi i\dfrac{m}{\rho} Z_m(k_\rho\rho) + 
\hat{\mathbf{e}}_zi k_{z} Z_m(k_\rho\rho)
\right]e^{im\phi}e^{ik_zz}
\\
&\mathbf{M}_{m}\left(\vec r, \vec k\right)=\left[
\hat{\vec e}_\rho i\dfrac{m}{\rho}Z_m(k_\rho\rho) -
\hat{\vec e}_\phi \dfrac{\partial Z_m(k_\rho\rho)}{\partial \rho}
\right]e^{im\phi}e^{ik_zz}
\\
&\mathbf{N}_{m}\left(\vec r, \vec k\right)=\left[
\hat{\vec e}_\rho i\dfrac{k_z}{k_\rho}\dfrac{\partial Z_m(k_\rho\rho)}{\partial \rho} -
\hat{\vec e}_\phi \dfrac{k_z}{k_\rho}\dfrac{m}{\rho}Z_m(k_\rho\rho) +
\hat{\vec e}_z \dfrac{k_\rho^2}{k} Z_m(k_\rho\rho)
\right]e^{im\phi}e^{ik_zz}
\label{eq:cylindrical_vh}
\end{aligned}
\end{equation}

%%% Eigenfrequency equation %%%%%%%%%%%%%%%%%%%%%%%%%%%%%%%%%%%%%%%%%%%%%%%%%%%%%%%%%%%%%%%%%%%%%%
\section{Derivation of the eigenfrequency equation of the BICs in a spherical resonator}
\label{sec:eigs_sphere}
Because displacement field satisfies the Helmholtz equation \eqref{eq:helmholtz}, it can be expanded using vector harmonics in a following manner
\begin{subequations}
    \begin{align}
    &\vec u^\mathrm{in}(\vec r) = \sum_i a_i \vec M_i(\vec r, \vec k_s) + b_i \vec N_i(\vec r, \vec k_s) + c_i \vec L_i(\vec r, \vec k_p),
    \label{eq:expansion_inside}
    \\
    &\vec u^\mathrm{out}(\vec r) = \sum_i d_i \vec L_i^o(\vec r, \vec k_0),
    \label{eq:expansion_outside}
    \end{align}
    \label{eq:expansion}
\end{subequations}
where, superscripts $\mathrm{"in"}$, $\mathrm{"out"}$ reflect the belonging of the field to the domain inside or outside of the resonator, $i = \{\ell, m\}$ for spherical vector harmonics and $i = m$ for cylindrical vector harmonics, $\vec k_s$ and $\vec k_p$ are the wavevectors of the shear and pressure modes respectively inside the resonator and $\vec k_0$ is the wavevector of the pressure waves in the surrounding fluid. Since fluid supports only purely longitudinal pressure waves, no $\vec M$ and $\vec N$ harmonics are present in the expansion (\ref{eq:expansion_outside}) and index $"o"$ at the longitudinal harmonic in the same expansion  represents the fact that spherical Hankel $h^{I}_\ell(k r)$ or Hankel $H^{I}_m(k_\rho \rho)$ function is used to build the solution of the Helmholtz equation (equations (\ref{eq:solution_helmholtz_sph}) and (\ref{eq:solution_helmholtz_cyl})), while for the $\vec M(\vec r, \vec k_s)$, $\vec N(\vec r, \vec k_s)$ and $\vec L(\vec r, \vec k_p)$ harmonics in the expansion (\ref{eq:expansion_inside}) spherical Bessel $j_\ell(k r)$ or Bessel $J_m(k_\rho \rho)$ function is taken as $z_{\ell}(k r)$ or $Z_m(k_\rho\rho)$. This choice was made on the basis of the following physical considerations
\begin{enumerate}
    \item Field outside of the resonator is the outgoing wave $\Rightarrow$ $z_\ell(k r)\:\&\:Z_m(k_\rho \rho)\to h^I_\ell(k r)\:\&\:H^I_m(k_\rho \rho)$,
    \item Field inside of the resonator should be finite at the origin $\Rightarrow$ $z_\ell(k r)\:\&\:Z_m(k_\rho \rho)\to j_\ell(kr)\:\&\:J_m(k_\rho \rho)$.
\end{enumerate}
Using equation (\ref{eq:pu_connection}) one can rewrite expansion of the field outside of the resonator \eqref{eq:expansion_outside} in terms of the pressure field
\begin{equation}
    p^\mathrm{out}(\vec r) =\sum_i \rho_0 \omega^2 d_i \psi_i^o(\vec r. \vec k_0),
\end{equation}
For the solid-fluid interface boundary conditions are
\begin{equation}
     \hat{\vec \sigma}_{n\tau} = 0,  \:\:  \hat{\vec \sigma}_{nn} = -p^\mathrm{out},\:\: u_n^\mathrm{in} = u_n^\mathrm{out},
    \label{eq:boundary_conditions}
\end{equation}
where indices $n$ and $\tau$ denote components normal and tangential to the boundary respectively.

Consider a spherical acoustic resonator with radius $a$ made of isotropic solid material immersed in a fluid.
Due to the symmetry of the system, the spherical coordinate system will be the most convenient to use.
In the spherical system of coordinates, strain tensor is written as \cite{landauelasticity}
\begin{equation}
    \begin{aligned}
&\varepsilon_{r r}=\dfrac{\partial u_{r}}{\partial r},
&
&\varepsilon_{\theta \theta}=\dfrac{1}{r} \dfrac{\partial u_{\theta}}{\partial \theta}+\dfrac{u_{r}}{r},
\\
&\varepsilon_{\varphi \varphi}=\dfrac{1}{r \sin \theta} \dfrac{\partial u_{\varphi}}{\partial \varphi}+\operatorname{ctg} \theta \dfrac{u_{\theta}}{r}+\frac{u_{r}}{r},
&
&\varepsilon_{r \theta}=\dfrac{1}{2}\left[\dfrac{1}{r} \dfrac{\partial u_{r}}{\partial \theta}+\dfrac{\partial u_{\theta}}{\partial r}-\dfrac{u_{\theta}}{r}\right],
\\
&\varepsilon_{r \varphi}=\dfrac{1}{2}\left[\dfrac{1}{r \sin \theta} \dfrac{\partial u_{r}}{\partial \varphi}+\dfrac{\partial u_{\varphi}}{\partial r}-\dfrac{u_{\varphi}}{r}\right],
&
&\varepsilon_{\theta \varphi}=\dfrac{1}{2}\left[\dfrac{1}{r}\left(\dfrac{u_{\varphi}}{\partial \theta}-u_{\varphi} \operatorname{ctg} \theta\right)+\dfrac{1}{r \sin \theta} \dfrac{\partial u_{\theta}}{\partial \varphi}\right].
\end{aligned}
\label{eq:strain_sph}
\end{equation}
One can show that
\begin{equation*}
    2\eps_{r\th} = \dfrac{\partial u_\th}{\partial r}-\dfrac{u_\th}{r}+\dfrac{1}{r}\dfrac{\partial u_r}{\partial\th}
    = -\dfrac{1}{r}\left[r\dfrac{\partial u_\th}{\partial r}+u_\th\dfrac{\partial r}{\partial r}-\dfrac{\partial u_r}{\partial \th}\right]+2\dfrac{\partial u_\th}{\partial r}
    = -\left(\rot \vec u\right)_\ph+2\dfrac{\partial u_\th}{\partial r},
\end{equation*}
and
\begin{equation*}
   2\eps_{r\phi}=\dfrac{1}{r\sin\th}\dfrac{\partial u_r}{\partial \ph}+\dfrac{\partial u_\ph}{\partial r}-\dfrac{u_\ph}{r} = \\
    = \dfrac{1}{r}\left[\dfrac{1}{\sin\th}\dfrac{\partial u_r}{\partial 
    \phi}-r\dfrac{\partial u_\ph}{\partial r}-u_\ph\dfrac{\partial r}{\partial r}\right]+2\dfrac{\partial u_\ph}{\partial r}
    =\left(\rot \vec u\right)_\th+2\dfrac{\partial u_\ph}{\partial r}.
\end{equation*}
Using equations \eqref{eq:hooke} and \eqref{eq:strain_sph} one can write boundary conditions \eqref{eq:boundary_conditions} for for the boundary between solid sphere and fluid environment
\stepcounter{equation}
\begin{enumerate}
    \item $\sigma_{r\th}=0$
    \begin{equation}
        \left(\rot \vec u^\iin\right)_\ph-2\dfrac{\partial u_\th^\iin}{\partial r}=0,
        \tag{\theequation a}
        \label{eq:boundary_sph_rth}
    \end{equation}
    \item $\sigma_{r\ph}=0$
    \begin{equation}
       \left(\rot \vec u^\iin\right)_\th+2\dfrac{\partial u_\ph^\iin}{\partial r}=0,
       \tag{\theequation b}
    \end{equation}
    \item $\sigma_{rr}=0$
\begin{equation}
    \lambda \div\vec u^{\iin}+2\mu\dfrac{\partial u_r^\iin}{\partial r}+p_\out=0,
    \tag{\theequation c}
\end{equation}
\item $u_n^\iin = u_n^\out$
\begin{equation}
    u_r^\iin=u_r^\out.
    \tag{\theequation d}
    \label{eq:boundary_sph_u}
\end{equation}
\end{enumerate}

Substituting expansion \eqref{eq:expansion} into the boundary conditions \eqref{eq:boundary_sph_rth} - \eqref{eq:boundary_sph_u} one will obtain set of equations for coefficients $a_{\ell m}$, $b_{\ell m}$, $c_{\ell m}$ and $d_{\ell m}$. Due to the orthogonality of the vector spherical harmonics, equations \eqref{eq:boundary_sph_rth} - \eqref{eq:boundary_sph_u} should be satisfied for every $\ell$ and $m$ separately. The equations \eqref{eq:boundary_sph_rth} - \eqref{eq:boundary_sph_u} can be presented in the matrix form
\begin{equation}
\left.
\underbrace{
    \begin{bmatrix}
    {k_s} N_\ph - 2{\partial_r}M_\th & {k_s}M_\ph-2\partial_rN_\th & -2\partial_r L_\th & 0 
    \\
    {k_s} N_\th + 2{\partial_r}M_\ph & {k_s}M_\th+2{\partial_r}N_\ph & 2{\partial_r}L_\ph & 0 
    \\
    0 & 2\mu \partial_rN_r & \lambda\operatorname{div}\vec L+2\mu \partial_r L_r & \rho_0\omega^2\psi^o
    \\
    0 & N_r & L_r & -L_r^o
    \end{bmatrix}
    }_{\hat{\boldsymbol{\mathrm{D}}}_{\ell m}}
    \underbrace{
    \begin{bmatrix}
    a \\ b \\ c \\ d
    \end{bmatrix}
    }_{\vec f_{\ell m}}
    \right|_{r=a}
    =0,
    \label{eq:sphericmatrix}
\end{equation}
where indices $\ell$, $m$ and arguments of the functions are omitted to improve readability.
In order to bring the matrix $\hat{\boldsymbol{\mathrm{D}}}_{\ell m}$ to a more noble form, let's take a look at its first two rows. Explicit view of these rows is
\begin{equation}
    \begin{bmatrix}
    \dfrac{i m r }{\sin\th}\dfrac{\partial}{\partial r}\left(\dfrac{j_\ell^s}{r}\right)P_\ell^m
    &
    \dfrac{1}{k_s r^2}\dfrac{\partial P_\ell^m}{\partial \th}\left[\left(2 - k_s^2 r^2\right) j_\ell^s+2r\dfrac{\partial }{\partial r}\left(r\dfrac{\partial j_\ell^s}{\partial r}\right)\right]
    &
    2 \dfrac{\partial}{\partial r}\left(\dfrac{j_\ell^p}{r}\right)\dfrac{\partial P_\ell^m}{\partial\th}
    &
    0
    \\[0.2em]
    \\
    r\dfrac{\partial}{\partial r}\left(\dfrac{j_\ell^s}{r}\right)\dfrac{\partial P_\ell^m}{\partial \th}
    &
    \dfrac{-im}{k_s r^2}\dfrac{P_\ell^m}{\sin\th}\left[\left(2 - k_s^2 r^2\right) j_\ell^s+2r\dfrac{\partial }{\partial r}\left(r\dfrac{\partial j_\ell^s}{\partial r}\right)\right]
    &
    2\dfrac{-im}{\sin\th}\dfrac{\partial}{\partial r}\left(\dfrac{j_\ell^p}{r}\right)P_\ell^m
    &
    0
    \end{bmatrix} e^{im\ph},
    \label{eq:rows_matrix_sph}
\end{equation}
where arguments of $j_\ell(k r)$ and $P_\ell^m(\cos\th)$ are omitted for the sake of brevity and indices $s$ and $p$ at the $j_\ell^{s,p}$ coincide with those of the wavevectors $k_{s,p}$ in the argument of the spherical Bessel function.
By doing operations
\begin{subequations}
    \begin{align}
        &\boldsymbol \kappa_1 +  \boldsymbol \kappa_2\dfrac{\sin\th}{im}\dfrac{\partial_\th P_\ell^m}{P_\ell^m} = \boldsymbol\kappa_1^\prime,
        \\
        &\boldsymbol\kappa_2 -\boldsymbol\kappa_1^\prime \dfrac{m}{i\sin\th}\dfrac{P_\ell^m\partial_\th P_\ell^m}{m^2\csc^2\th \left(P_\ell^m\right)^2 - \left(\partial_\th P_\ell^m\right)^2} = \boldsymbol\kappa_2^\prime,
    \end{align}
\end{subequations}
where $\boldsymbol\kappa_1$ and $\boldsymbol\kappa_2$ are the rows of the matrix \eqref{eq:rows_matrix_sph}, one can make the matrix $\hat{\boldsymbol{\mathrm{D}}}_{\ell m}$ block-diagonal, and the solution of the matrix equation \eqref{eq:sphericmatrix} will be
\begin{equation}
\operatorname{det}\hat{\boldsymbol{\mathrm{D}}}_{\ell m} =
    \underbrace{\operatorname{det} \hat{ \mathbf{D}}_{\ell m}^{1 \times 1}}_{\text {BIC }} \cdot \underbrace{\operatorname{det} \hat{ \mathbf{D}}_{\ell m}^{3 \times 3}}_{\text {Rad. modes }}=0.
\end{equation}
Therefore, matrix equation \eqref{eq:sphericmatrix} has two nontrivial solutions
\begin{equation}
    \operatorname{det} \hat{\mathbf{D}}_{\ell m}^{1 \times 1}=0,\:\:\operatorname{det} \hat{\mathbf{D}}_{\ell m}^{3 \times 3}=0,
\end{equation}
where solution
\begin{equation}
   \operatorname{det} \hat{\mathbf{D}}_{\ell m}^{1 \times 1}= -i\dfrac{r \sin\th}{m  P_\ell^m(\cos\th)}\left[ \dfrac{m^2}{\sin^2\th}\left(P_\ell^m(\cos\th)\right)^2 - \left(\dfrac{\partial P_\ell^m(\cos\th)}{\partial\th}\right)^2 \right]\dfrac{\partial}{\partial r}\left(\dfrac{ j_\ell(k_s r)}{r}\right)e^{i m \phi}=0
\end{equation}
is satisfies the matrix equation \eqref{eq:sphericmatrix} for $b=c=d=0$ thus giving eigenfrequencies for BICs and can be rewritten as
\begin{equation}
    (1-\ell)j_\ell(x)+x j_{\ell+1}(x) = 0,
    \label{eq:eigenfrequency_m1}
\end{equation}
where $x = k_s a$. The roots of Eq.~\eqref{eq:eigenfrequency_m1} are shown in Table~\ref{tab:m_eigenfrequency}.

%table 1 %%%%%%%%%%%%%%%%%%%%%%%%%%%%%%%%%%%
\begin{table}[h]
\caption{\label{tab:m_eigenfrequency}
First 3 roots of Eq.~\eqref{eq:eigenfrequency_m1} for BICs of the spherical resonator for $\ell = 1,2,3,4$.
}
\begin{ruledtabular}
\begin{tabular}{ccccc}
$\ell$&
\textrm{1}&
\textrm{2}&
\textrm{3}&
\textrm{4}
\\
\colrule
$x_1$ & 5.76 & 2.50 & 3.86 & 5.09 \\
$x_2$ & 9.10 & 7.13 & 8.44 & 10.51 \\
$x_3$ & 12.32 & 10.51 & 11.88 & 13.21\\
\end{tabular}
\end{ruledtabular}
\end{table}
%%%%%%%%%%%%%%%%%%%%%%%%%%%%%%%%%%%%%%%%%%%%%%

%%%%%% BIC In Cylinder %%%%%%%%%%%%%%%%%%%%%%%%%%%%%%%%%%%%%%%%%%%%%%%%%%%%%%%%%%%%%%%%%%%%%%%%%%%%%
\section{Acoustic bound states in the continuum in a solid cylindrical resonator}
\label{sec:eigs_cyl}
Since displacement field of BIC is completely localized in the resonator, eigenmode problem for BICs can be solved using stress-free boundary conditions
\begin{equation}
     \hat{\vec \sigma}_{n\tau} = 0,  \:\:  \hat{\vec \sigma}_{nn} = 0,
    \label{eq:boundary_conditions_free}
\end{equation}
In cylindrical system of coordinates components of the Cauchy stress tensor are presented as \cite{landauelasticity}
\begin{equation}
    \begin{aligned}
        &\eps_{\r \r}=\dfrac{\partial u_{\r}}{\partial \r}, 
        &
        &\eps_{\varphi \varphi}=\dfrac{1}{\r} \dfrac{\partial u_{\varphi}}{\partial \varphi}+\dfrac{u_{\r}}{\r},
        &
        &\eps_{z z}=\dfrac{\partial u_{z}}{\partial z},
        \\
        &\eps_{\varphi z}=\dfrac{1}{2}\left[\dfrac{1}{\r} \dfrac{\partial u_{z}}{\partial \varphi}+\dfrac{\partial u_{\varphi}}{\partial z}\right],
        &
        &\eps_{\r z}=\dfrac{1}{2}\left[\dfrac{\partial u_{\r}}{\partial z}+\dfrac{\partial u_{z}}{\partial \r}\right],
       &
        &\eps_{\r \varphi}=\dfrac{1}{2}\left[\dfrac{\partial u_{\varphi}}{\partial \r}-\dfrac{u_{\varphi}}{\r}+\dfrac{1}{\r} \dfrac{\partial u_{\r}}{\partial \varphi}\right],
    \end{aligned}
\end{equation}
where non-diagonal components can be reduced to the following form
\begin{equation*}
    2\eps_{\r\ph}=\dfrac{1}{\r}\left(-\r\dfrac{\partial u_\ph}{\partial \r}-u_\ph\dfrac{\partial \r}{\partial \r}+\dfrac{\partial u_\r}{\partial\ph}\right)+2\dfrac{\partial u_\ph}{\partial \r}=-\left(\rot \vec u\right)_z+2\dfrac{\partial u_\ph}{\partial \r},
\end{equation*}
\begin{equation*}
    2\eps_{\r z}=\dfrac{\partial u_\rho}{\partial z} - \dfrac{\partial u_z}{\partial \rho} + 2\dfrac{\partial u_z}{\partial \rho} =  \left(\rot\vec u \right)_\ph+2\dfrac{\partial u_z}{\partial \r}.
\end{equation*}
Hereby explicit form of boundary conditions is
\stepcounter{equation}
\begin{itemize}
    \item At the side surface of the cylinder - for $\rho = a$
    \begin{enumerate}
    \item $\sigma_{\r\phi}=0$
    \begin{equation}
            -\left(\rot \vec u\right)_z+2\dfrac{\partial u_\ph}{\partial \r}=0,
            \tag{\theequation a}
            \label{eq:cyl_boundary_1}
    \end{equation}
    
    \item $\sigma_{\r z}=0$
    \begin{equation}
            \left(\rot\vec u\right)_\ph+2\dfrac{\partial u_z}{\partial \r}=0,
            \tag{\theequation b}
    \end{equation}
    
    \item 
    $\sigma_{\r\r}=0$
    \begin{equation}
        \lambda \div \vec u +2\mu\dfrac{\partial u_\r}{\partial \r}=0,
        \tag{\theequation c}
    \end{equation}
    \end{enumerate}
    
\item On the bases of the cylinder - for $z = \pm h/2$
\stepcounter{equation}
\begin{enumerate}
    \item $\sigma_{z\ph}=0$
    \begin{equation}
           \left(\rot\vec u\right)_\r+2\dfrac{\partial u_\ph}{\partial z}=0,
            \tag{\theequation a}
    \end{equation}
     \item  $\sigma_{z\r}=0$ 
    \begin{equation}
            \left(\rot\vec u\right)_\ph+2\dfrac{\partial u_z}{\partial \r}=0,
            \tag{\theequation a}
    \end{equation}
    \item $\sigma_{zz}=0$
    \begin{equation}
        \lambda \div \vec u + 2\mu \dfrac{\partial u_z}{\partial z}=0,
    \tag{\theequation c}
    \label{eq:cyl_boundary_2}
    \end{equation}
\end{enumerate}
\end{itemize}
where $a$ and $h$ are the radius and height of the cylindrical resonator, respectively.
Before proceeding, it is important to make a few notes that will simplify the mathematical calculations.
\begin{enumerate}
    \item Due to the symmetry of the cylinder only modes with azimuthal number $m = 0$ will be the BICs \cite{gladyshev2020symmetry}. This can be also understood from the explicit view of the $\vec M_m$ vector cylindrical harmonic \eqref{eq:cylindrical_vh}, since in has nonzero radial component for $m\neq 0$.
    \item Vector cylindrical harmonics generated using $\psi(\vec r) = \psi(\rho,\phi)e^{ik_z z}$ will not satisfy boundary conditions for the bases of the cylinder since $e^{ix}\neq 0$ for $x \in \mathbb R$, therefore, basis of the standing waves should be used
    \begin{equation}
        \psi_m(\vec r, \vec k) = J_m(k_\rho \rho)\left[A^o\sin(k_z z) + A^e\cos(k_z z)\right]e^{im\phi},
    \end{equation}
    where indices $"o"$ and $"e"$ near the amplitudes $A^o$ and $A^e$ reflect the odd or even nature of the sine and cosine relative to zero.
    \item All harmonics with different azimuthal number $m$ and parity $q = o,e$ are fully decoupled and therefore can be considered separately.
\end{enumerate}
Considering all the comments above, the decomposition of the displacement field is written as
\begin{equation}
         \vec u^q_0 (\vec r) = a^q \vec M_0^q(\vec r, \vec k_s) + b^q\vec N_0^q(\vec r, \vec k_s) + c^q\vec L_0^q(\vec r, \vec k_p).
    \label{eq:expansion_cyl}
\end{equation}
By direct substitution of the expansion \eqref{eq:expansion_cyl} into the boundary conditions \eqref{eq:cyl_boundary_1} - \eqref{eq:cyl_boundary_2} one can obtain two matrix equations
\begin{equation}
    \left.
    \underbrace{
        \begin{bmatrix}
            2\partial_\rho M_\phi - k_s N_z
            &
            0
            &
            0
            \\
            0
            &
            2\partial_\rho N_z + k_s M_\phi
            &
            2\partial_\rho L_z
            \\
            0
            &
            2\mu \partial_\rho N_\rho
            &
            \lambda \Delta \psi_p + 2\mu\partial_\rho L_\rho
        \end{bmatrix} }_{\hat{\vec D}_m^{\rho}}
        \underbrace{
        \begin{bmatrix}
            a
            \\
            b
            \\
            c
        \end{bmatrix}}_{\vec f_m^\rho}
        \right|_{\rho = a}
        = 0
        \label{eq:matrix_cyl_r}
\end{equation}
for the side surface of the cylinder and
\begin{equation}
\left.
    \underbrace{
    \begin{bmatrix}
            2\partial_z M_\phi + k_s N_\rho
            &
            0
            &
            0
            \\
            0
            &
            2\partial_\rho N_z + k_s M_\phi
            &
            2\partial_\rho L_z
            \\
            0
            &
            2\mu\partial_z N_z
            &
            \lambda \Delta\psi_p + 2\mu \partial_z L_z
    \end{bmatrix}}_{\hat{\vec D}_m^z}
    \underbrace{
    \begin{bmatrix}
    a
    \\
    b
    \\
    c
    \end{bmatrix}}_{\vec f_m^z}
\right|_{z=\pm h/2}
    = 0
    \label{eq:matrix_cyl_z}
\end{equation}
for the bases of the cylinder. Indices $0$, $q$ and arguments of the functions in the equations \eqref{eq:matrix_cyl_r} and \eqref{eq:matrix_cyl_z} are omitted for the sake of brevity. One of solutions of matrix equations \eqref{eq:matrix_cyl_r} and \eqref{eq:matrix_cyl_z}, which corresponds to the BIC is
\begin{equation}
    \left\{
    \begin{aligned}
        &2\partial_\rho M_\phi - k_s N_z \Big |_{r = a} = 0,
        \\
        &2\partial_z M_\phi + k_s N_\rho \Big |_{z = \pm h/2} = 0.
    \end{aligned}
    \right.
    \label{eq:nonf_eigenfreq_cyl}
\end{equation}
Explicitly, the equations \eqref{eq:nonf_eigenfreq_cyl} can be written as
\begin{equation}
    \begin{aligned}
    &\underbrace{
        \left\{
        \begin{aligned}
            &J_2(k_\rho a) = 0,
            \\
            &\cos\left( k_z h/2 \right) = 0,
        \end{aligned}
        \right.
    }_{\text{odd mode}}
    & 
    &\underbrace{
        \left\{
        \begin{aligned}
            &J_2(k_\rho a) = 0,
            \\
            &\sin\left( k_z h/2 \right) = 0,
        \end{aligned}
        \right.
    }_{\text{even mode}}
    \end{aligned}
\end{equation}
thus giving $k_{\rho}$ and $k_z$ components of the wavevector of the odd or even BICs. Eigenfrequencies of the BICs can be calculated as
\begin{equation}
    \omega^2 = c_t^2 \left[\left(\dfrac{\alpha_n}{a}\right)^2 + \left(\dfrac{\pi q}{h}\right)^2\right],
\end{equation}
where $\alpha_n$ is the $n$'th root of the Bessel function of the second order $J_2(x)$, and $q = \pm1, \pm3, \pm5 ...$ for the odd and $0, \pm2, \pm4...$ for the even modes.

\section{Classification of the modes in solids of $D_{\infty h}$ symmetry}
\label{sec:the_big_table}
% table 2 %%%%%%%%%%%%%%%%%%%%%%%%%%%%%%%%%%%
\begin{table}[h]
\caption{\label{tab:quadrupole_BIC1}  Multipole content and classification of eigenmodes, and selection rules for eigenmode excitation by a plane wave in the objects of $D_{\infty h}$ symmetry (spheroid, cylinder, symmetrical dimer etc).    
}
\begin{ruledtabular}
\begin{tabular}{cclcclccc}
Irrep&
\begin{tabular}[c]{@{}c@{}}Degree of\\ degeneracy \end{tabular}&
Mode&
\begin{tabular}[c]{@{}c@{}}Azimuthal\\ number \end{tabular}&
Parity&
\text{Multipoles}&
\begin{tabular}[c]{@{}c@{}}Excitation\\ along $x$ or $y$ \end{tabular}&
\begin{tabular}[c]{@{}c@{}}Excitation\\ along $z$ \end{tabular}
&
\begin{tabular}[c]{@{}c@{}}Excitation\\at oblique \\ incidence \end{tabular}
\\
\colrule
$A_{1u}$ & 1 & \begin{tabular}[c]{@{}l@{}}Radiative\\ mode  \end{tabular} & $m=0$ & odd & 
\begin{tabular}[c]{@{}l@{}}
$\mathbf{L}_{1,0}$,...,$\mathbf{L}_{2s+1,0}$\\
$\mathbf{N}_{1,0}$,...,$\mathbf{N}_{2s+1,0}$ \\
\end{tabular}
& No & Yes & Yes\\
\colrule
$A_{1g}$ & 1 & \begin{tabular}[c]{@{}l@{}}Radiative\\ mode  \end{tabular} & $m=0$ & even & 
\begin{tabular}[c]{@{}l@{}}
$\mathbf{L}_{0,0}$,...,$\mathbf{L}_{2s,0}$\\
$\mathbf{N}_{2,0}$,...,$\mathbf{N}_{2s,0}$ \\
\end{tabular}
& No & Yes & Yes\\
\colrule
$A_{2g}$ & 1 & BIC & $m=0$ & even & $\mathbf{M}_{1,0}$,...,$\mathbf{M}_{2s+1,0}$ & No & No & No\\
\colrule
$A_{2u}$ & 1 & BIC & $m=0$ & odd & $\mathbf{M}_{2,0}$,...,$\mathbf{M}_{2s,0}$ & No & No & No \\
\colrule
$E_{1g}$ & 2 & \begin{tabular}[c]{@{}l@{}}Radiative\\ mode  \end{tabular} & $m=\pm1$ & odd & \begin{tabular}[c]{@{}l@{}}$\mathbf{L}_{2,\pm1}$...,$\mathbf{L}_{2s,\pm1}$\\ $\mathbf{N}_{2,\pm1}$...,$\mathbf{N}_{2s,\pm1}$ \\ 
$\mathbf{M}_{1,\pm1}$...,$\mathbf{M}_{2s+1,\pm1}$
\end{tabular}  & No & No & Yes\\
\colrule
$E_{1u}$ & 2 & \begin{tabular}[c]{@{}l@{}}Radiative\\ mode  \end{tabular} & $m=\pm1$ & even & \begin{tabular}[c]{@{}l@{}}$\mathbf{L}_{1,\pm1}$...,$\mathbf{L}_{2s+1,\pm1}$\\ $\mathbf{N}_{1,\pm1}$...,$\mathbf{N}_{2s+1,\pm1}$ \\ 
$\mathbf{M}_{2,\pm1}$...,$\mathbf{M}_{2s,\pm1}$
\end{tabular}  & Yes & No & Yes\\
\colrule
$E_{2u}$ & 2 & \begin{tabular}[c]{@{}l@{}}Radiative\\ mode  \end{tabular} & $m=\pm2$ & odd & \begin{tabular}[c]{@{}l@{}}
$\mathbf{L}_{3,\pm2}$,...,$\mathbf{L}_{2s+1,\pm2}$\\
$\mathbf{N}_{3,\pm2}$,...,$\mathbf{N}_{2s+1,\pm2}$ \\
$\mathbf{M}_{2,\pm2}$,...,$\mathbf{M}_{2s,\pm2}$
\end{tabular} & No & No & Yes \\
\colrule
$E_{2g}$ & 2 & \begin{tabular}[c]{@{}l@{}}Radiative\\ mode  \end{tabular} & $m=\pm2$ & even & \begin{tabular}[c]{@{}l@{}}
$\mathbf{L}_{2,\pm2}$,...,$\mathbf{L}_{2s,\pm2}$\\
$\mathbf{N}_{2,\pm2}$,...,$\mathbf{N}_{2s,\pm2}$ \\
$\mathbf{M}_{3,\pm2}$,...,$\mathbf{M}_{2s+1,\pm2}$
\end{tabular} & Yes & No & Yes \\
\colrule
$E_{3g}$ & 2 & \begin{tabular}[c]{@{}l@{}}Radiative\\ mode  \end{tabular} & $m=\pm3$ & odd & \begin{tabular}[c]{@{}l@{}}
$\mathbf{L}_{4,\pm3}$,...,$\mathbf{L}_{2s,\pm3}$\\
$\mathbf{N}_{4,\pm3}$,...,$\mathbf{N}_{2s,\pm3}$ \\
$\mathbf{M}_{3,\pm3}$,...,$\mathbf{M}_{2s+1,\pm3}$
\end{tabular} & No & No & Yes \\
\colrule
$E_{3u}$ & 2 & \begin{tabular}[c]{@{}l@{}}Radiative\\ mode  \end{tabular} & $m=\pm3$ & even & \begin{tabular}[c]{@{}l@{}}
$\mathbf{L}_{3,\pm3}$,...,$\mathbf{L}_{2s+1,\pm3}$\\
$\mathbf{N}_{3,\pm3}$,...,$\mathbf{N}_{2s+1,\pm3}$ \\
$\mathbf{M}_{4,\pm3}$,...,$\mathbf{M}_{2s,\pm3}$
\end{tabular} & Yes & No & Yes \\
\colrule
\vdots & \vdots  & \vdots & \vdots  & \vdots  & \vdots  & \vdots  & \vdots  & \vdots
\end{tabular}
\end{ruledtabular}
\end{table}
%%%%%%%%%%%%%%%%%%%%%%%%%%%%%%%%%%%%%%%%%%%%%%
Table~\ref{tab:quadrupole_BIC1} shows the classification of eigenmodes in solids of $D_{\infty h}$ symmetry (spheroid, cylinder, symmetrical dimer etc). We assume that the rotation axis is the $z$-axis. As the considered bodies of revolution have a mirror symmetry with respect to the $xy$-plane, we can define a parity of the modes under transformation $z\rightarrow-z$ as follows    
\begin{equation}
\left[\begin{array}{c}
u_{\rho}(\rho, \varphi,-z) \\
u_{\varphi}(\rho, \varphi,-z) \\
u_{z}(\rho, \varphi,-z)
\end{array}\right]=\sigma_{z}\left[\begin{array}{r}
u_{\rho}(\rho, \varphi, z) \\
u_{\varphi}(\rho, \varphi, z) \\
-u_{z}(\rho, \varphi, z)
\end{array}\right].
\end{equation}
If $\sigma_z=1$ then the mode is defined as even, and if $\sigma_z=-1$ then the mode is defined as odd.

\end{document}